\documentclass[preprintnumbers,amsmath,11pt,amssymb,floatfix,superscriptaddress,nofootinbib]{article}

\def\mytitle#1{\setcounter{equation}{0}
\setcounter{footnote}{0}
\begin{flushleft}\Large\textbf{#1}\end{flushleft}
\vspace{0.25cm}}
\def\myname#1{\leftline{{\large #1}}\vspace{-0.13cm}}
\def\myplace#1#2{\small\begin{flushleft}\textit{#1}\\
\texttt{#2}\end{flushleft}}

\def\myclassification#1{\small\noindent
Keywords :
       #1\vspace{0.5cm}}
\usepackage{graphicx}% Include figure files
\begin{document}

\mytitle{Quantum Corrected Schwarzschild Black Hole: Inner Horizon Thermodynamic Behaviors}

\myname{$Abhijit~ Mandal^{*}$\footnote{abhijitmandal.math@gmail.com} and $Ritabrata~
Biswas^{\dag}$\footnote{biswas.ritabrata@gmail.com}}
\myplace{*Department of Mathematics, Jadavpur University, Kolkata $-$ 700 032, India.\\$\dag$ Department of Mathematics, Bankura University, Bankura $-$ 722 146, India.}{} 
 
%%%%%%%%%%%%%%%%%%%%%%%%%%%%%%%%%%%%%%%%%%%%%%%%%%%%%%%%%%%%%%%%%%%%
\begin{abstract}
The thermodynamic properties of Cauchy horizon is a matter of interest. Event horizon and Cauchy horizon together can enlighten us about the micro-states of a black hole. In addition, if we consider a black hole metric modified with quantum terms, which is not forcing the geodesics to focus at a singularity, the study of horizons becomes much more interesting. The spacelike behavior inside the Cauchy horizon has a deep impact on the related thermodynamics. We analyze different thermodynamic product to check whether a right left string theory mode's addition type representation for the concerned thermodynamic parameters is possible or not. Stability of Cauchy horizon is studied.
\end{abstract}
%%%%%%%%%%%%%%%%%%%%%%%%%%%%%%%%%%%%%%%%%%%%%%%%%%%%%%%%%%%%%%%%%%%%

\myclassification{Black hole physics; Quantum aspects of black holes, evaporation, thermodynamics; Relativity and gravitation.}

%%%%%%%%%%%%%%%%%%%%%%%%%%%%%%%%%%%%%%%%%%%%%%%%%%%%%%%%%%%%%%%%%%%%%%%%
\section{Introduction}

The Bogoliubov transformations \footnote{The Bogoliubov transformation coefficients, $\alpha_{jk}$ and $\beta_{jk}^*$-s follow:
$$\sum_k\left(\alpha_{jk}\alpha_{j'k}^* - \beta_{jk}\beta_{j'k}^*\right) = \delta_{jj'}~~~,$$ such that any two sets of creation and annihilation operators can be presented as $$a_j = \sum_k\left(\alpha^*_{jk}b_k - \beta^*_{jk}b^{\dag}_k\right)~~and ~~b_k = \sum_j\left(\alpha_{jk}a_j + \beta^*_{jk}a^{\dag}_j\right).$$
	Mean number of particles created into mode $k$ is $$\left\langle N_k \right\rangle = _{in}\left\langle0|b^{\dag}_{k}b_k|0\right\rangle_{in} = \sum_j |\beta_{jk}|^2 .$$ 
	Now, $\left\langle N_k \right\rangle\neq 0$ only when any of $\beta_{jk}\neq 0$, i.e., when mixing of positive and negative frequency solutions occurs.} of creation and annihilation operators lead to the result that if there is any mixing of positive and negative frequency solutions, then particles are created by a gravitational field\cite{Hawking1}.
	If Schwarzschild black hole(BH hereafter) of mass $M$ is been considered as the central engine of the gravitational field, the mean number of particles created into mode $\omega lm$ is given by 
	$$N_{\omega lm} = \sum_{\omega'}|\beta_{\omega' \omega lm}|^2=\frac{1}{e^{8 \pi M \omega - 1}},~~~~~\beta=Bogoliubov~~ transformation ~~coefficient.$$ 
	This is a Plank spectrum with a temperature, $$T= \frac{1}{8 \pi M},$$
	this is called the Hawking temperature of a Schwarzschild BH. As the temperature is inversely proportional to the mass of the BH, the specific heat of the BH is negative. Once the BH starts to evaporate, there is no way to halt it  and from the semi-classical point of view, the end result is complete annihilation of the BH.
	Now, we consider a quasi-static process during which a stationary BH of mass $M$, angular momentum $J$ and surface area $A$ is taken to a new stationary BH with parameters $M + \delta M$, $J + \delta J$ and $ A + \delta A $. This perturbation is done by a small quantity of matter described by the stress-energy tensor $T_{\alpha \beta}$.
	So, the mass and angular momentum of the BH increase by amounts:
	$$\delta M = - \int_H T^\alpha_\beta \xi^\beta_{(t)} dH_\alpha ~~ and
	~~\delta J = - \int_H T^\alpha_\beta \xi^\beta_{(\phi)} dH_\alpha ,$$
	where $\xi^{\beta}_{(t)}$ is the Killing vector for the translation and $\xi^\beta_{(\phi)}$ is the rotational Killing vector. The integration is over the entire event horizon $H$ and $dH_\alpha = - \xi_\alpha dA dv$, is the surface element on $H$, with co-ordinates $y = (v, \theta^{A})$. $\xi^\alpha = \xi^{\alpha}_{(t)} + \Omega_H \xi^\alpha_{(\phi)}$, where $\Omega_H$ is the angular velocity of the BH.
	All these definations brings us to,
	$$\delta M - \Omega_H \delta J = \int_H T_{\alpha \beta} \left(\xi^\beta_{(t)} + \Omega_H \xi^\beta_{(\phi)}\right) \xi^\alpha dA ~dv = \int dv \oint_H T_{\alpha\beta} \xi^{\alpha} \xi^{\beta} dA.$$
	Now a simplified Raychoudhuri's equation \cite{Poisson} for terms like expansion $\theta$ and shear $\sigma_{\alpha \beta}$ are of first order in $T_{\alpha \beta}$. Hence neglecting all quadratic term, we can write  
	$$\frac{d\theta}{dv} = \kappa \theta - 8 \pi T_{\alpha \beta} \xi^\alpha \xi^\beta.$$
Using the fact that both before and after the perturbation, the BH is stationary, i.e., $\theta(v = \pm \infty) = 0$, we obtain,
$$\delta M - \Omega_H \delta J = - \frac{1}{8 \pi} \int dv \oint_H \left(\frac{d\theta}{d v} - \kappa \theta\right) dA = \frac{\kappa}{8 \pi} \int dv \oint_H \theta dA.$$
	Now using the defination of $\theta$ as the fractional rate of change of the congruence's cross-sectional area, we get
$$\delta M - \Omega_H \delta J =\frac{\kappa}{8 \pi} \int dv \oint_H \left(\frac{1}{dA} \frac{d}{dv} dA\right) dA =\frac{\kappa}{8 \pi} \oint_H dA =\frac{\kappa}{8 \pi} \delta A,$$

which is the statement of first law of BH thermodynamics.

The area theorem established by Hawking\cite{Hawking2} gives rise to the second law of BH mechanics which states that if the null energy condition is satisfied, the surface area of the BH can never decrease. We relate the entropy of a BH to one quarter of the area of their event horizon in Planck units. A physical realization  of the entropy can be made if we consider thermal radiation and follow what entropy it would have when it forms a BH. In Planck units $G =c = \hbar = \kappa = 1$,  a ball of radiation at temperature $T$ and radius $R$ has mass $M \sim T^4 R^3$, and entropy $S \sim T^3R^3.$ The radiation will form a BH when $R \sim M$ which implies $T \sim M^{-\frac{1}{2}}$ and hence $S \sim M^\frac{3}{2}$. So, the entropy of the resulting BH is $S_{BH} \sim M^2$. So any BH with $M>>1$, i.e., mass much larger than the Planck mass, has an entropy much larger than the entropy of the thermal radiation that formed it. The entropy should be a measure of the number of the underlying quantum states.
	Now, the explanation of BH entropy requires the knowledge of three facts of string theory\cite{Green}. The first says whenever one quantizes a string in flat spacetime, there is an infinite tower of massive states. For every integer $N$ there are states with $M^2 \sim N/l^2_s$, where $l_s$ is a new length scale set by the string tension. The number of string states at excitation level $N>>1$ is $e^{S_s}$ where,
	$$S_s \sim \sqrt{N},$$
	i.e., the string entropy is proportional to the mass in string units.
	The second fact is that the string interactions are governed by a string coupling constant $`g'$, determined by a scalar field called dilaton. In $4D$ Newton's gravitational constant $G$ is related to $g$ and the string length $l_s$ by $G \sim g^2 l_s^2$. 
	The third fact is that the classical spacetime metric is well defined in string theory only when the curvature is less than the string scale $\frac{1}{l_s^2}$.
	One characteristic string model property is to establish entropy as the sum of contributions from left and right moving excitations of the strings. Now, for a BH there is only one state for which the thermodynamic variables are defined, i.e., the event horizon. This happens only when the BH is characterised by its mass(eg. Schwarzschild BH). But except the mass, whenever we introduce other parameters, the scenario changes. Say, for Reissner Nordstr$\ddot{o}$m BH, if the charge of the BH is less than its mass(measured in geometric units $G=c=1$), then two horizons exist. Two or more horizons are at least mathematically possible for many other BHs\cite{Pradhan}.
	Standard thermodynamic variables, defined at the outer event horizon, are mirrored by an independent set of thermodynamic variables, defined at the inner event horizon or Cauchy horizon. The left- and right-moving thermodynamics of the string theory correspond to the sum and the difference of the outer and the inner horizon thermodynamics of BH. the relation was established by direct inspection for large classes of extremal and near extremal BHs\cite{Horowitz1, Halyo, Horowitz2}. 
	In $5D$ the entropy of a general rotating BH \cite{Cvetic1}
	$$S= 2 \pi \left[\sqrt{\frac{1}{4}\mu^3\left(\prod^3_{i=1} cosh \delta_i + sinh \delta_i\right)^2 - J^2_L} + \sqrt{\frac{1}{4}\mu^3\left(\prod^3_{i=1} cosh \delta_i - sinh \delta_i\right)^2 - J^2_R}\right]. $$
	This form of the entropy may be interpreted as an indication that it derives from two independent microscopic contributions, and each of these may be attributed to a gas of strings\cite{Cvetic1, Cvetic2, Larsen1}. If we be able to describe the entropy as this summation, it indicates that the interactions between two kinds of modes must be weak. The corrosponding levels of the effective string can be written as\cite{Cvetic3}
	$$N_L =\frac{1}{4}\mu^3\left(\prod^3_{i=1} cosh \delta_i + sinh \delta_i\right)^2 - J^2_L ~~~ and 
	~~~N_R = \frac{1}{4}\mu^3\left(\prod^3_{i=1} cosh \delta_i - sinh \delta_i\right)^2 - J^2_R.$$
	So, for large levels,
	$$S=S_L + S_R = 2\pi \left(\sqrt{N_L} + \sqrt{N_R}\right).$$
	
	The microscopic degrees of freedom of the BH are described in terms of those of a conformal field theory - without gravity - living in the boundary. This construction gave the inner $(A_-)$ and outer $(A_+)$ Killing horizons of BHs obeying the product formula 
  $$\frac{A_+ A_-}{(8\pi G_3)^2} = N_{R} - N_{L}~~,$$
where $N_{R}$ and $N_{L}$ are number of right and left moving excitations of the $2D$ conformal field theory. Generalizing for any asymptotically flat BH in $d$ - spacetime dimensions \cite{Larsen1}, 
$$\frac{A_+ A_-}{(8\pi G_d)^2} \in Z~~, ~the ~set~of~integers~. $$
In a nut shell, this says the product of areas is independent of BH mass and solely depends on the quantized charges. This suggestive general relation for BHs encouraged purely gravitational investigation of the product of area of BHs \cite{Cvetic4}. For Kerr-Newman BH area - product is been investigated \cite{Ansorg}.

	Contrary to the outer horizon thermodynamics, it is not very clear where the inner - horizon has any relevance for a statistical accounting of the BH entropy. First law for the inner horizon is schematically  
\begin{equation}\label{inner_horizon}
-dM = T_{-}\frac{dA_{-}}{4G_{5}} - (\Omega^{-} dJ + \phi^{-}_{E} dQ + \phi^{-}_{m} dq)~~.
\end{equation}	
	The reference \cite{Curir} has studied the above relation for Kerr BHs and references \cite{Cvetic3} studied the same for more generalized BH. For the sake of comparison the first law for the outer horizon is 
\begin{equation}\label{outer_horizon}
dM = T_{+}\frac{dA_{+}}{4G_{5}} + (\Omega^{+} dJ + \phi^{+}_{E} dQ + \phi^{+}_{m} dq).
\end{equation}	
	The minus signs in (\ref{inner_horizon}) are due to the Killing horizon vector field being spacelike inside the BH event horizon. Keeping similarity to the negative energies within ergosphere, negative energy ($-M$) is assigned to the inner horizon as well. For the perturbatively unstable nature of the inner horizon one may follow \cite{Marlof, Castro1}. 
				If someone considers a number of different higher curvature modifications of the Einstein Hilbert action, the entropy will not generally be proportional to horizon area and the product relations for the horizon areas and entropies will no longer be equivalent. According to \cite{Wald1}, BH entropy can be viewed as a Noether charge, it might be more natural to except that the entropy product formula, rather than the product of the areas would be the correct generalization for modification to Einstein gravity. There exist(s) such references like \cite{Castro2}, where considering the holding of entropy product formula as a `success', a try to find `failures' has been taken, i.e., where the product of the horizon entropy depends on the mass. For such `failures' Smarr relation is modified. Though there is no obvious relation between the horizon entropy products and Smarr relations, but for these failing cases both are drastically modified by the presence of higher curvature terms.
Quantum study towards gravitation and strongly gravitating compact objects leads us to find the quantum modified BH metric.

%%%%%%%%%%%%%%%%%%%%%%%%%%%%%%%%%%%%%%%%%%%%%%%%%%%%%%%%%%%%%%%%%%%%%%%%%%%%

The Raychoudhuri's equation can be modified by replacing  the classical trajectories by quantal trajectories\cite{Das}. The quantum Raychoudhuri equation has been found to prevent focusing of geodesics, and hence prevents the formation of singularities. Quantum Raychoudhuri equation for null geodesics are found in \cite{Das, Ali} and a  modified Schwarzschild metric is derived: 
\begin{equation}\label{metric}
ds^2 = -(1 - \frac{2M}{r} + \frac{\hbar\eta}{r^2}) dt^2 + \frac{1}{(1 - \frac{2M}{r} + \frac{\hbar\eta}{r^2})} dr^2 + r^2 d \Omega^2,
\end{equation}

where $\hbar \rightarrow 0$ and $\eta$ is a constant. The constant $\eta$ must be dimensionless because $\hbar$ has dimensions of $(length)^2$ in geometric units.

While studying the particle trajectory towards the event horizon and Cauchy horizon interesting phenomena are found. In physically realistic solutions, a signal approaching $r_-$ gets infinitely blue shifted. This makes the Cauchy horizon unstable\cite{Chandrasekhar, Poisson2}. Again \cite{Poisson3} shows that, a perturbation, the inner horizon of a charged BH becomes a singularity with infinite spacetime curvature. This leads us to believe that the Cauchy horizon is a true physical singularity. For the modified Schwarzschild metric\cite{Ali2} has found a  Cauchy horizon of Planckian size. Now the immediate question arises that what should be the thermodynamic behavior of such a non-point like singularity. We motivated ourselves to analyze the specific heat, free energy and the other parameters. Even it will be interesting to observe where the thermodynamic parameters of these two horizons are interacting(i.e., their product can be represented as the function of the mass) or they can be represented in a way that they resemble some quantum state(free of mass).

In the next section we will calculate different thermodynamic products for quantum modified Schwarzschild metric and physically analyze the thermodynamic behavior of event horizon and Cauchy horizon. At the end we will go for a brief conclusion of the work.

%%%%%%%%%%%%%%%%%%%%%%%%%%%%%%%%%%%%%%%%%%%%%%%%%%%%%%%%%%%%%%%
\section{Thermodynamic Product Study for Quantum Schwarzschild BH}
The interesting observation regarding this work was the formation of an extremal BH which yields a frozen horizon and hence the BH evaporation stops forming a BH remnant. The physical singularity of the BH turned to be a timelike  singularity, i.e., the quantal trajectories will never go there and will not feel the singularity at all. 

From the mertic(\ref{metric}) we get the form of the BH horizons,
\begin{equation}
r_\pm = M \pm \sqrt{M^2 - \hbar \eta }~.
\end{equation}
Here, $r_+$ corresponds to event horizon and $r_{-}$ corresponds to Cauchy horizon.
Product of these two is
\begin{equation}
r_{+}r_-=\hbar \eta ~~,
\end{equation}
clearly stating the fact that this quantity is global.
The Hawking temperature, from the first law of thermodynamics will take the form 
\begin{equation}
T_{\pm}=\frac{1}{4 \pi} \lim_{r_{\pm} \to 0} \frac{d g_{tt}}{d r}, 
\end{equation}
And we have,
\begin{equation}
T_+ T_- = \pm \frac{\sqrt{M^2 - \hbar \eta}}{2 \pi (M \pm \sqrt{M^2 - \hbar \eta})}.
\end{equation}
So this is depending on mass of BH and hence is not a global property.

Usually entropy of a BH calculated from the so-called area formula and which equals to one-quarter of the horizon area. In higher derivative curvature terms, in general, the entropy of a BH doesnot match with the area formula. But as a thermodynamic system BH must obey the first law of thermodynamics $dM = T dS$ \cite{Cai1}. Integrating the first law, we have,
$$S = \int T dM$$
and $$S_\pm = \int_0^{r_\pm} T^{-1} \frac{\delta M}{\delta r} dr$$
where we have imposed the physical assumption that entropy vanishes when the horizons of BH shrinks.
Now, entropy becomes
\begin{equation}
S_{\pm} = \pm 2 \pi r_{\pm},~~~~~~~\Rightarrow~~S_+S_- = - 4 \pi^2 \hbar \eta.
\end{equation}
which is a BH mass free global quantity and stating that the interactions may be weak. The entropy($S_-$) for the radius of Cauchy horizon is negative! 
There are different points of view about this situation. One can assume that above condition is just the equation to remove the non-physical domain of theory parameters. However, it is difficult to justify such proposal. On another side, one can conjecture that classical thermodynamics is not applied here and negative entropy simply indicates to new type of instability. Such a situation is justified \cite{Cvetic5}. 

The free energy will take the form
\begin{equation}
F_{\pm}=M-T_{\pm}S_{\pm}= M - \frac{\sqrt{M^2 - \hbar \eta}}{(M \pm \sqrt{M^2 - \hbar \eta})}
\end{equation}
We are not writting the expression of $F_+F_-$ but it is depending on M saying that this quantity is again not universal.
Now proceeding towards heat capacity, we find 
$$C_{\pm}=T_{\pm}\frac{\partial S_{\pm}}{\partial T_{\pm}}=\frac{\partial M}{\partial S_{\pm}}\frac{\partial S_{\pm}}{\partial T_{\pm}}=\frac{\partial M}{\partial T_{\pm}}=\frac{1}{\frac{\partial T_{\pm}}{\partial M}}$$
\begin{equation}
 = \frac{2 \pi \left[ \hbar \eta \mp 2 M \left(\pm M +\sqrt{M^2 - \hbar \eta}\right)\right]}{2 \mp \frac{M}{\sqrt{M^2 - \hbar \eta}}}
\end{equation}

So the product is 
\begin{equation}
C_+C_- = \frac{(2 \pi \hbar \eta)^2 (M^2 - \hbar \eta)}{3M^2 - 4 \hbar \eta }
\end{equation}

\begin{figure}[h]
\begin{center}

~~~~~~~~~~Fig.1a~~~~~~~~~~~~~~~~~~~~~~~~~~~Fig.1b~~~~~\\
\includegraphics[height=2.5in, width=1.5in]{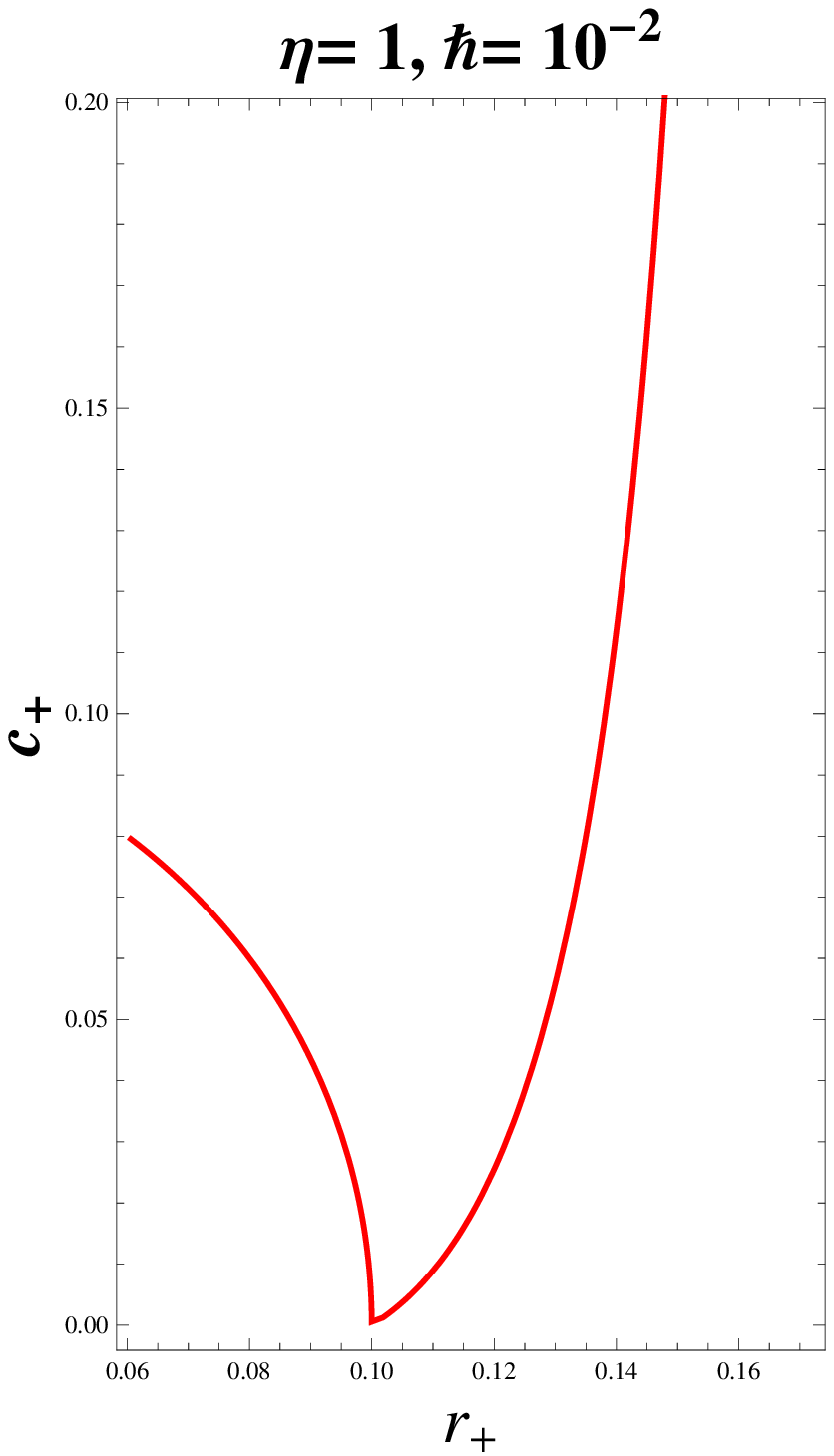}~~~~
\vspace{.6cm}
\includegraphics[height=2.5in, width=1.5in]{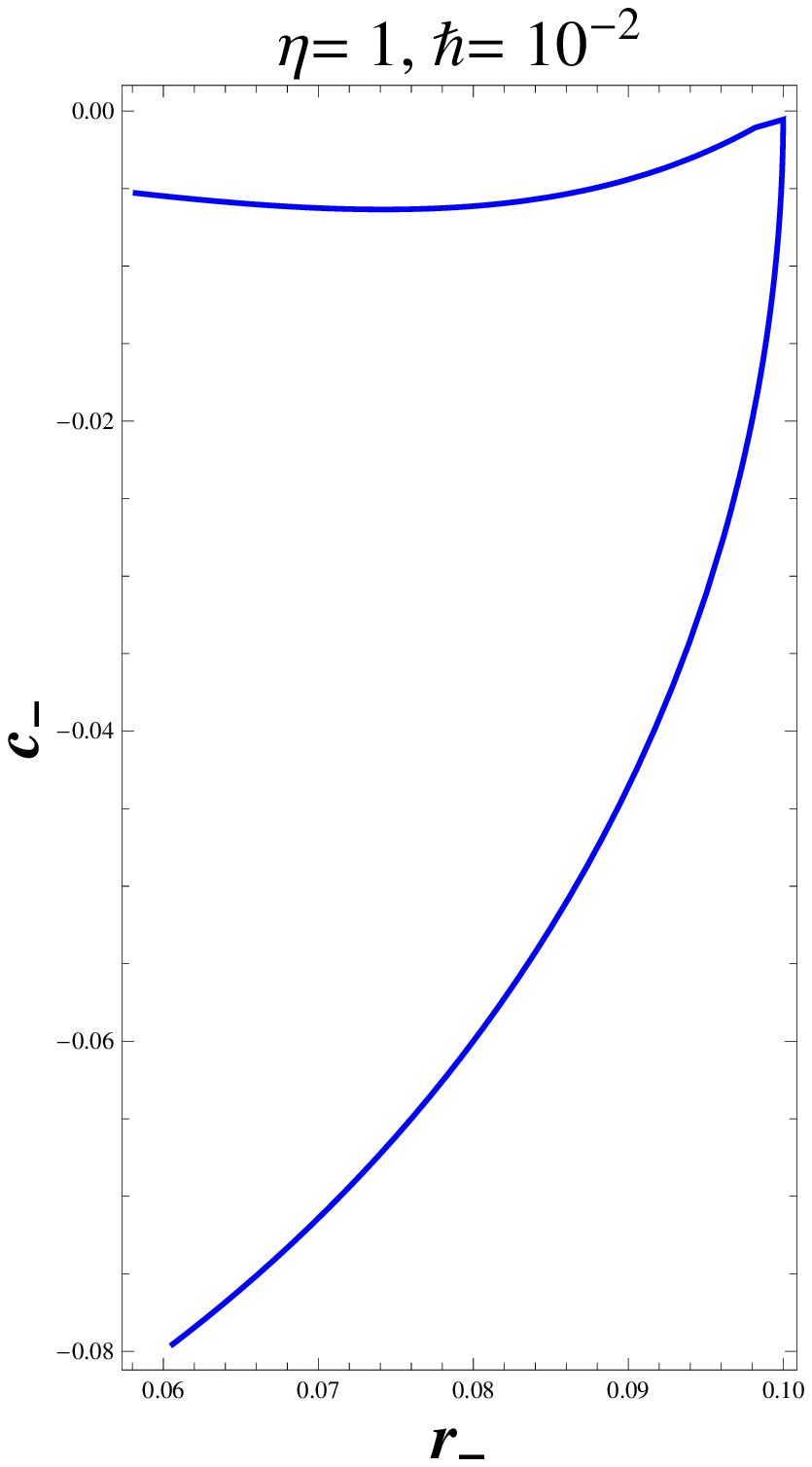}~~\\
Fig $1a$ represent the variation of specific heat($C_+$) for event horizon with respect to $r_+$ .\\
Fig $1b$ represent the variation of specific heat($C_-$) for Cauchy horizon with respect to $r_-$ .

\end{center}
\end{figure}

In figure $1a$ we analyze the specific heat($C_+$) for event horizon, which is plotted with respect to the radius of  event horizon($r_+$) where $\eta =1$ and $\hbar=10^{-2}$. In this figure, when the value of $r_+$ is small, we find a phase in which the $C_+$ vs $r_+$ curve is slowly decreasing and at last becomes $C_+ = 0$ for a certain value of $r_+$(say, $r^*$). On the other hand for the large value of $r_+$ we find another phase which starts from the value $C_+ =0$ at $r_+ = r^*$ and the curve is strictly increasing. So, both the phases have a common point at the line $C_+ =0$ at $r_+= r^*$. The first phase radiates energy/heat to decrease the temperature but the second phase, on the contrary, requires more and more energy/heat for the same increment in temperature. Here overall the sign of $C_+$ is positive and we will find one and only phase for each value of $r_+$. The all over positive $C_+$ indicates stable phases of the spacetime confined in the event horizon.

	In figure $1b$ we plot specific heat($C_-$) for Cauchy horizon with respect to the radius of Cauchy horizon($r_-$), where $\eta =1$ and $\hbar=10^{-2}$. We observe two phases in $C_-$ vs $r_-$ graph and for each value of $r_-$ we will get two values of $C_-$. In one phase, the $C_-$ vs $r_-$ curve is slowly increasing and finally takes the value $C_- =0$ at $r_-= r^*$. On the other hand, the curve is strictly increasing for the second phase and finally it will also takes the value $C_- =0$ at $r_-= r^*$. So like figure $1a$, here also both the phases has a common point at the line $C_- =0$ at $r_-= r^*$. Here overall the sign of $C_-$ is negative which indicates unstable phases of the space time confined in the Cauchy horizon. But for $r_- > r^*$, we donot see any physical value of $C_-$. It seems the unstable Cauchy horizon of different phases reaches $r^*$, where the specific heat($C_-$) for Cauchy horizon vanishes, i.e., how much energy will be required for unit change of temperature can not be predicted there and may be the Cauchy horizon at all do not exist after that.

\begin{figure}[h]
\begin{center}

~~~~~~~~~~Fig.2a~~~~~~~~~~~~~~~~~~~~~~~~~~~Fig.2b~~~~~\\
\includegraphics[height=2.5in, width=1.5in]{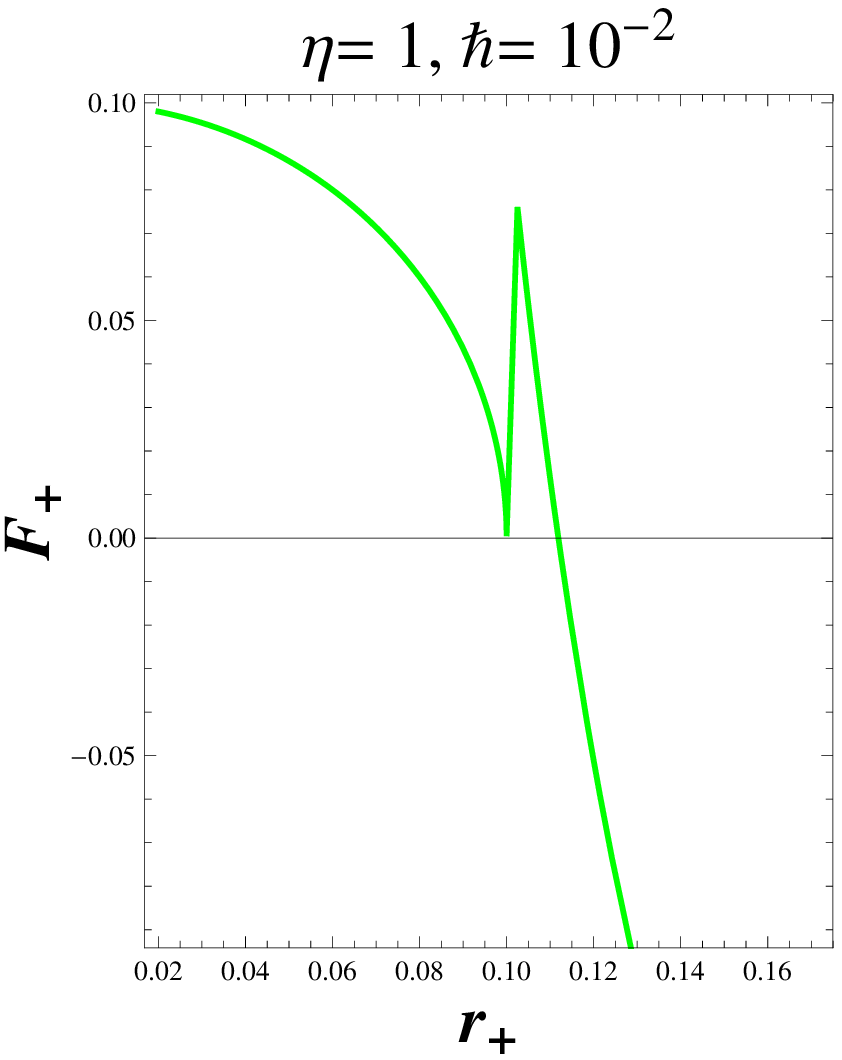}~~~~
\vspace{.6cm}
\includegraphics[height=2.5in, width=1.5in]{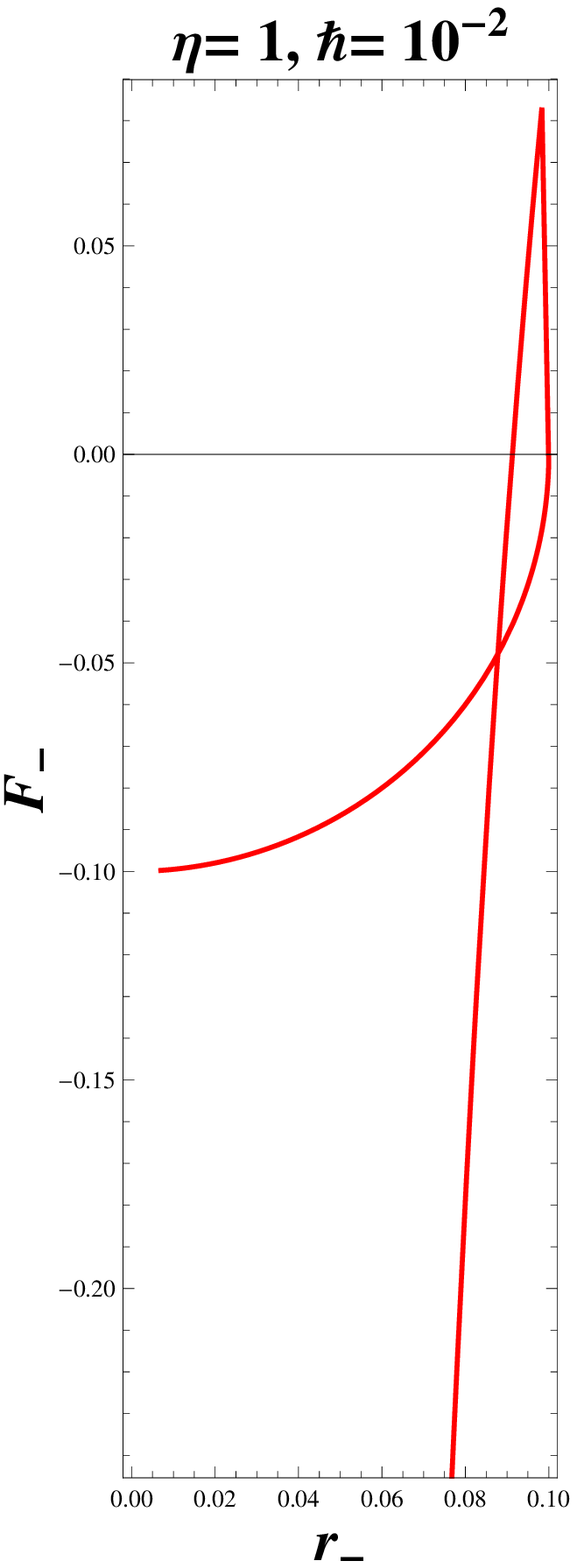}~~\\
Fig $2a$ represent the variation of free energy($F_+$) for event horizon with respect to $r_+$ .\\
Fig $2b$ represent the variation of free energy($F_-$) for Cauchy horizon with respect to $r_-$ .

\end{center}
\end{figure}

Now, the graph $2a$ shows the variation of the free energy($F_+$) for event horizon against the radius of event horizon($r_+$). In this figure, we observe two phases, one for small values of $r_+$ and another for large values of $r_+$. The $F_+$ vs $r_+$ curve for both the phases are decreasing. The first phase ends at $r_+= r^*$ and the second phase starts at $r_+= r^*$. We will find a finite energy jump at $r_+= r^*$. Since, the specific heat($C_+$) for event horizon vanishes at $r_+= r^*$, may be there is a huge amount of energy absorption happens which is the reason behind this energy jump at $r_+= r^*$.

In figure $2b$ we show the free energy($F_-$) for Cauchy horizon against the radius of Cauchy horizon($r_-$), where $\eta =1$ and $\hbar=10^{-2}$. We observe two phases in $F_-$ vs $r_-$ curve. In the first phase the $C_+$ vs $r_+$ curve is slowly increasing. In this phase the overall value of $F_-$ is negative except the point $r_- =r^*$. On the another side the second phase is strictly increasing and both the phases have two common points, one of them is $r_- =r^*$ and another is $r^{**}$(, say) where $r^{**}<r^*$. But for $r_- > r^*$, we donot see any physical value of $F_-$ which may be indicates that the Cauchy horizon at all do not exist after that.
%%%%%%%%%%%%%%%%%%%%%%%%%%%%%%%%%%%%%%%%%%%%%%%%%%%%%%%%%%%%%%
\section{Conclusion}
A new semiclassical approach for quantum gravity by replacing classical trajectories with quantal trajectories, i.e., by obtaining a quantum modified version of Raychoudhuri equation is found in literature\cite{Das}. This equation is found to prevent the focussing of geodesics and hence prevent the formation of singularities. While studied for cosmological purpose, it was observed that it is possible to resolve big bang singularity using quantal geodesics\cite{Ali}. For the BH solution\cite{Ali2}, it is found that the freely falling particles are similarly behaving as they do for Reissner-Nordstr$\ddot{o}$m BH (the mathematical similarity arises at $Q^2\rightarrow \hbar \eta$).
The spacetime confined within event horizon is always stable. We found two different phases when analyzing the specific heat and the free energy against the radius horizon. Here, at a certain value of the radius of event horizon these phases meet each other for both specific heat and free energy.

We observe that only the BH entropy product is universal quantity, whereas the Hawking temperature product, free energy product and specific heat product are not universal quantities because they all are depends on mass parameter. The input of quantum corrections says that the left and right modes are interacting and may not very weakly.

Between the outer horizon $r_+$ and the inner Cauchy horizon($r_-$), the co-ordinates $r$ becomes timelike and a falling particle must continue inwards. According to \cite{Ali2} $below$ $r= r_-$ the co-ordinate $r$ is spacelike and the falling particle can move in the direction of increasing $r$ until $r=r_-$ where $r$ becomes timelike again but in the reverse direction. Even we found two different phases by analyzing the specific heat and the free energy at the Cauchy horizon. This two phases exist for the same value of $r$ though and at a particlar radius of Cauchy horizon they meet each other. At this common cuspidal type meeting point of these two phases(in $C_-$ vs $r_-$ graph) the specific heat is zero. This means even not absorbing or radiating, the BH can increase its temperature at this particular critical radii. Once $r>r_-$ we find no value/ physical curve of $C_-$, which indicating the fact that the Cauchy horizon for such a big quantum BH even does not exist at all. For the quantum corrected metric, the singularities are supposed to be timelike. So the singularities can not be felt. So, somehow our result supports this fact.

We want to end this article with some interesting comments. Inspired by the Penrose diagram analysis\cite{Misner, Carroll}  the authors of \cite{Ali2} have speculated that, a infalling particle gets into a wormhole at the cauchy horizon and gets out of a white hole in a different universe. The two phases shown in $1b$ and $2b$ may be indicating the phases but in different universes. These two meets at some particular radius of Cauchy horizon of the concerned universe's BH.

%%%%%%%%%%%%%%%%%%%%%%%%%%%%%%%%%%%%%%%%%%%%%%%%%%%
\vspace{.1 in}
{\bf Acknowledgement:}
RB thanks Inter University Center for Astronomy and Astrophysics(IUCAA), Pune, India for Visiting Associateship. Authors thank IUCAA for local hospitality. This work was done during a visit there. Authors thank Prof. Subenoy Chakraborty, Department of Mathematics, Jadavpur University for fruitful discussions.
%%%%%%%%%%%%%%%%%%%%%%%%%%%%%%%%%%%%%%%%%%%%%%%%%%%%%%%%%%%%%

\end{document}